\documentclass[aip,graphicx]{revtex4-1}
\usepackage[cp1251]{inputenc}
\usepackage[T2A]{fontenc}
\usepackage[english]{babel}
\usepackage{amssymb,latexsym,amsmath,amscd}
\usepackage{graphicx,color,framed}
\begin{document}
\selectlanguage{english}
\title{Metal-Organic Framework Breathing in Electric Field: A Theoretical Study}
\author{\firstname{Andrei L.} \surname{Kolesnikov}}
\email[]{kolesnikov@inc.uni-leipzig.de}
\affiliation{Institut f\"{u}r Nichtklassische Chemie e.V., Permoserstr. 15, 04318 Leipzig, Germany}

\author{\firstname{Yury A.} \surname{Budkov}}
\email[]{ybudkov@hse.ru}
\affiliation{Tikhonov Moscow Institute of Electronics and Mathematics, National Research University Higher School of Economics, Tallinskaya st. 34, 123458 Moscow, Russia}
\affiliation{G.A. Krestov Institute of Solution Chemistry of the Russian Academy of Sciences, Akademicheskaya st. 1, 153045 Ivanovo, Russia}

\author{\firstname{Jens} \surname{M\"{o}llmer}}
\affiliation{Institut f\"{u}r Nichtklassische Chemie e.V., Permoserstr. 15, 04318 Leipzig, Germany}

\author{\firstname{Michael G.} \surname{Kiselev}}
\affiliation{G.A. Krestov Institute of Solution Chemistry of the Russian Academy of Sciences, Akademicheskaya st. 1, 153045 Ivanovo, Russia}

\author{\firstname{Roger} \surname{Gl\"{a}ser }}
\affiliation{Institut f\"{u}r Nichtklassische Chemie e.V., Permoserstr. 15, 04318 Leipzig, Germany}

\begin{abstract}
In this manuscript, we study the electrically induced breathing of Metal-Organic Framework (MOF) within a 2D lattice model. The Helmholtz free energy of the MOF in electric field consists of two parts: the electrostatic energy of the dielectric body in the external electric field and elastic energy of the framework. The first contribution is calculated from the first principles of statistical mechanics with an account of MOF symmetry. By minimizing the obtained free energy and solving the resulting system of equations, we obtain the local electric field and the parameter of the unit cell (angle $\alpha$). The paper also studies the cross-section area of the unit cell and the polarization as functions of the external electric field. We obtain the hysteresis in the region of the structural transition of the framework. Our results are in qualitative agreement with the literature data of the molecular dynamics (MD) simulation of MIL-53(Cr).
\end{abstract}
\maketitle
\section{Introduction}
Stimuli-responsive materials attract the attention of researchers due to their numerous applications and remarkable properties \cite{Ge.2012,Bajpai.2008}. Among external stimuli, the most frequently used are adsorption/absorption\cite{Kitaura.2003,Kitagawa.2005,Ferey.2009,Schneemann.2014}, temperature\cite{Kitaura.2003,Henke.2013}, electric and magnetic fields \cite{Tam.2017,Ghoufi.2017,Meng.2010}, mechanical stimuli \cite{Zhang.2017,Neimark.2011,Coudert.2013}, light \cite{Meng.2010} etc. There are numerous types of materials that respond to certain environmental changes.  Among them are polymers \cite{Bawa.2009,Chae.2015,Gurovich.1994,Gurovich.1995,Tsori.2009,Kolesnikov.2017,Brilliantov.2016,Brilliantov.2016, Brilliantov.2017,Budkov.2015,Budkov.2016,Budkov.2018} (and as a consequence derived materials and composites containing polymers), liquid crystals, ionic liquids, solid adsorbents, etc. It is well known that solid materials can deform during gas adsorption with typical magnitudes of volume strain less than one percent \cite{Gor.2017,Kolesnikov.2018}. However, there are groups of materials, for example, aerogels, whose deformation can exceed tens of percent \cite{Reichenauer.2001}. Another interesting class of materials with an enhanced flexible response is soft metal-organic frameworks (MOFs). MOFs are inorganic-organic hybrids, with a lattice structure consisting of metal ions bound by ligands \cite{Chang.2015}. Although the mechanism of flexibility is different from that of polymers and depends on the type of MOF \cite{Kitaura.2003,Kitagawa.2005,Ferey.2009,Schneemann.2014,Kobalz.2016,Kobalz.2016b}, they react to the similar stimuli, among the most frequently used is gas adsorption.
Due to the MOF unique properties, there are a lot of potential applications, for example, gas separation \cite{Hamon.2009, Gu.2010, Maes.2010, Hahnel.2016}, gas/heat storage \cite{Jeremias.2012, Henninger.2010, Murray.2009, Tagliabue.2011, BastosNeto.2012} sensing \cite{Achmann.2009, Lu.2010, Sapchenko.2011} controllable capture and release \cite{Thallapally.2012, Luo.2014, Mason.2015}, {\sl etc}. The unusual properties of  MOFs, especially, Gate-Opening-MOFs, attract a lot of attention of researchers and motivate them to find new possibilities to control material properties contact-free.

The recent literature shows intensive work in the field of stimuli-responsive MOFs, and here we will discuss some of these publications. Sievers et al. \cite{Sievers2013} studied the dependence of metal-organic framework mesoparticles on water vapor pressure. The remarkable feature is that its crystalline structure consists of a two-dimensional coordination polymer, packed in parallel sheets, with organized clusters of water molecules lying between the sheets and bridging them via a dense H-bond network\cite{Sievers2013}. Authors observed that the particles respond with shrinking to the decrease of vapor pressure. Li et al. \cite{Li2016} presented dual stimuli-responsive MOF (magnetic PCN-250), which shows responses to both magnetic induction and ultraviolet (UV) light. Authors demonstrated that the combination of both triggers results in high $CO_2$ desorption at 1 bar. Huang et al. \cite{Huang2016} investigated the structural behavior of the gallium-based metal-organic framework having the MIL-53 topology. The dry material can switch between a narrow-pore phase and a large-pore phase by means of a temperature increase; also, the hysteresis accompanies the structural transition. Baimpos et al. \cite{Baimpos2015} studied real-time deformation of HKUST - 1 crystals caused by humidity adsorption and desorption. Also, the authors observed the nonmonotonic structural changes during the initial hydration of crystals. Namasani\cite{Namsani2018} et al., using density functional theory and molecular dynamics, studied the behavior of metal–organic frameworks in an external electric field. In particular, they showed the possibility to rotate the organic linkers with permanent dipole moment by changing the strength of the electric field. Tam et al. \cite{Tam.2017} demonstrated the potential mechanism of electric field controlled molecular gates. They are complex molecules with permanent dipole anchored on the host MOF and rotating by the changing of the direction of the electric field.
Recently, it has been shown that an external electric field has a great potential in separation processes \cite{Knebel.2017,Knebel.2018}, as the authors showed an increase in the separation factor for the mixture $C_3H_6/C_3H_8$. Moreover, by means of molecular dynamics (MD) simulations, Ghoufi and co-authors \cite{Ghoufi.2017} showed the reversible structural transition of $MIL-53(Cr)$ by applying an external electric field. In an empty host material, the first order transition with hysteresis was observed at 1 - 2 $V/nm$. It should be noted that electric field values are far above the breakdown electric field of air.  Schmid proposed in his work \cite{Schmid.2017} a possible molecular explanation of the electric field-induced phase transition. The mechanism is based on the mutual dipole-dipole interactions while the external electric field induces the average dipole moment of the interacting groups. In addition, the author proposed several possible mechanisms of polarization.

\begin{figure} [h]
\center{\includegraphics[scale = 1.2]{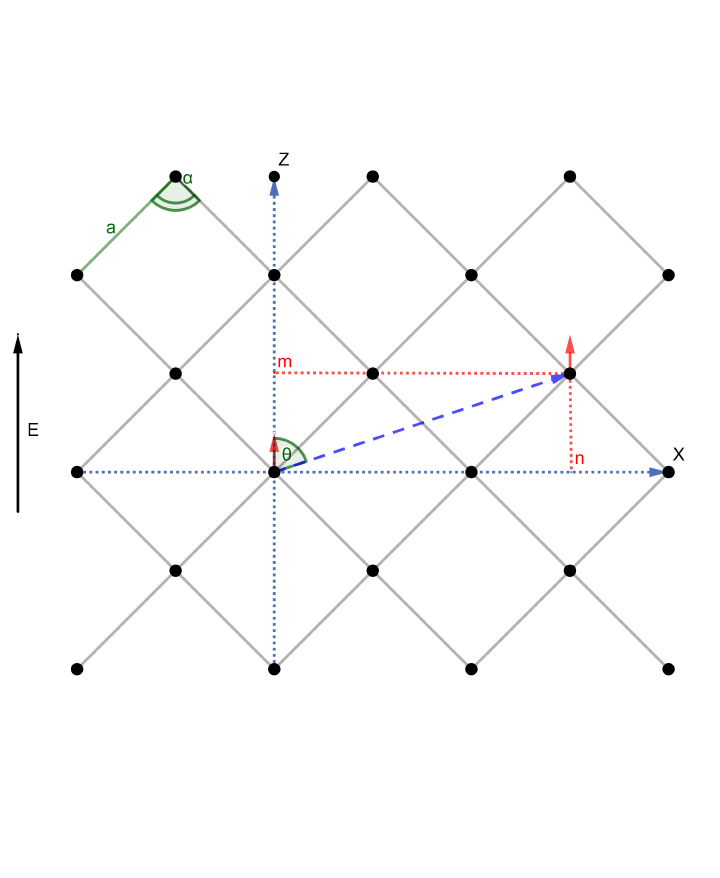}}
\caption{Illustration of the 2D MOF matrix, where E is the external electric field, $\theta$ is the sharp angle between the field direction and the dipole-dipole radius vector, a is the length of the linker. Each node represents the metal ion and the gray connections are the linkers. Red arrows are the induced dipoles on the two arbitrary chosen nodes, other dipoles are not shown for the simplicity reasons. The $\alpha$ is the expansion angle in the elastic Helmholtz free energy.}
\label{fig:0}
\end{figure}

To the best of our knowledge, there are no statistical models describing the effect of an external electric field on the MOFs. Thus, in this manuscript, we propose a model describing the phase transition of a solid matrix induced by dipole - dipole interactions under an external electric field. As a starting point, we use the physical model, proposed in the mentioned work \cite{Schmid.2017}, i.e. we consider a rhombic lattice 2D structure (see Fig. \ref{fig:0} ) with induced dipoles in the nodes of the lattice. In this approximation, each nod corresponds to the metal ion, whereas the organic linkers are treated with a fixed length $a$.


\section{Theory}
In the manuscript, we will follow the idea that the MOF transition is induced by mutual dipole-dipole interactions\cite{Schmid.2017}. Accordingly, the system is polarized by an external electric field and carry some polarization density. The possible polarization mechanisms are proposed in reference \cite{Schmid.2017}. The author supposes that the external electric field could reorient the $O-H$ groups, which carry local dipole moments. Also, as a positively charged metal ion is surrounded by negatively charged atoms, a strong electric field could deform the configuration and thus induce a dipole moment. The result of both mechanisms is the appearance of the local dipole moment induced by the electric field. Clearly, the list is not comprehensive, and highly sensitive to the structure of the MOF. As far as both induced dipoles are located in the close vicinity of an ion, for the seek of simplicity, we will put them in the same position - namely in the center of an ion. Thus, nodes can be treated as the positions of the induced local dipoles. Now, let us consider the model of MOF as a 2D lattice with $N$ junctions (see fig. 1). Each node represents the metal ion and the gray connections are the linkers.  Also, induced dipole moment can be decomposed into two contributions -- an orientational part from the permanent dipole moments and an instantaneous induced dipole moment from the molecular polarization. We assume, for simplicity, that both permanent and instantaneous induced dipole moments have only two possible opposite orientations along which the homogeneous constant electric field $E$ is applied, as it is shown in Fig. 1. The total Helmholtz free energy of MOF can be written as a sum of two terms:
\begin{equation}
F =F_{m}+F_{el},
\end{equation}
where the first term on the right hand side is the elastic free energy of the 2D matrix, which can be expressed as a function of angle ($\alpha$) and the second term is the electrostatic free energy of the lattice.   In order to obtain the elastic free energy, we expand it in the power-law series around equilibrium angle ($\alpha_0$) at zero electric field (the derivation is presented in the Supporting Information: Appentdix II):
\begin{equation}
F_{m}(\alpha) = F_{m,0}+\kappa_0 (\alpha - \alpha_0) + \kappa_1 (\alpha - \alpha_0)^2 + \kappa_2 (\alpha - \alpha_0)^3 + \kappa_3 (\alpha - \alpha_0)^4, \label{eq:elastic_energy}
\end{equation}
where $F_{m,0}$ is the free energy of MOF at $\alpha_0$ and the $\kappa_n$ ($n=1,2,3$) are the adjusted coefficients which describe the elastic behavior of the particular MOF. The first constant $\kappa_0$ is related to the initial stress in the MOF and will be discussed at the end of this section. The elastic coefficients $\kappa_1$, $\kappa_2$ and $\kappa_3$ were obtained, in order to qualitively reproduce the structural transition behaviour like in the MIL-53, namely the jump from one phase (lp-phase) to the 20other (np-phase). Naturally, the proposed elastic free energy does not describe the complex two stage transitions, which were obtained in the experiment \cite{Serre.2007}, but at the same time contains all the necessary information to describe the structural transition, driven by dipole-dipole interactions (see below).

In order to calculate the electrostatic contribution to the total free energy of MOF, we write the Hamiltonian of electrostatic interactions:
\begin{equation}
\label{Hamilt}
H_{el} =  \sum\limits_{i = 1}^{N} \left[ \frac{\xi_i^2}{2\gamma}  -pEs_i  - E\xi_i \right] + \frac{1}{8\pi \epsilon_0}\sum\limits_{i=1}^N\sum\limits_{j=1}^N\frac{(\xi_i + p s_{i})(\xi_j + p s_{j})}{ r_{ij}^3}\left(1-3\cos^2 \theta_{ij}\right),
\end{equation}
where $p$ is the permanent dipole moment, $\theta_{ij}$ is the angle between two dipoles, $\epsilon_0$ is the vacuum  permittivity, $s_i=\pm 1$ are the numbers specifying the projections of the permanent dipoles on the electric field direction, $\xi_i$ is the instantaneous induced dipole moments, and $\gamma$ is the molecular polarizability of nodes. The first sum in the Eq.(\ref{Hamilt}) takes into account the self energy of the instantaneous induced dipoles and the interaction between both permanent and induced dipoles with the external electric field. The second contribution takes into account dipole-dipole interactions between both types of dipoles.  Now, in accordance with the standard variation method, the Hamiltonian (\ref{Hamilt}) can be rewritten in terms of Hamiltonian $H_0$ of the reference system and the perturbation Hamiltonian $\Delta H$:
\begin{equation}
H_{el} = H_0 + \Delta H,
\end{equation}
\begin{equation}
H_0 = \sum\limits_{i = 1}^{N} \left[ \frac{\xi_i^2}{2\gamma}  -p\mathcal{E}s_i  - \mathcal{E}\xi_i \right]
\end{equation}
\begin{equation}
\Delta H = \sum\limits_{i=1}^N \left[ p s_i(\mathcal{E} - E) + \xi_i (\mathcal{E} - E)  \right] + \frac{1}{8\pi \epsilon_0}\sum\limits_{i=1}^N\sum\limits_{j=1}^N\frac{(\xi_i + p s_{i})(\xi_j + p s_{j})}{r_{ij}^3}\left(1-3\cos^2 \theta_{ij}\right),
\end{equation}
where we have introduced variational parameter $\mathcal{E}$, which physically describes the electrostatic mean field in the lattice. We would like to stress that we chose a lattice of non-interacting dipoles under $"$external$"$ electric field $\mathcal{E}$ as a reference system. Thus, the partition function $Z_0$\cite{Lu.2015,Budkov.2015b} of the reference system can be easily calculated, which yields
\begin{equation}
Z_0 = \sum\limits_{s_1 =\pm 1}\sum\limits_{s_2 =\pm 1}...\sum\limits_{s_N =\pm 1} \int\limits_{-\infty}^{\infty} \prod\limits_{i=1}^{N} d\xi_i e^{-\beta H_0}=\exp\left[-\beta F_0\right],
\end{equation}
where $\beta=1/k_{B}T$, $k_{B}$ is the Boltzmann constant, $T$ is the temperature. The reference Helmholtz free energy is
\begin{equation}
F_0 = -k_B T N \left[ \ln(e^{\beta \mathcal{E} p} + e^{-\beta \mathcal{E} p}) + \frac{1}{2}\ln(2\pi k_B T \gamma) + \frac{\gamma\mathcal{E}^2}{2 k_B T} \right].
\end{equation}
From the Bogolyubov inequality one can obtain the upper limit for the free energy:
\begin{equation}
F_{el} \leq F_0 + \left<\Delta H\right>,
\end{equation}
where the symbol $\left<..\right>$ denotes the average over statistics of the reference system:
\begin{equation}
\left<(..)\right> = \frac{1}{Z_0}\sum_{\{s_i\}} \int\limits_{-\infty}^{\infty} \prod\limits_{i=1}^{N} d\xi_i \exp\left[-\beta H_0\right](..).
\end{equation}
The average value of dipole parameter $s_i$ is
\begin{equation}
S=\left<s_i\right> = \frac{e^{\beta p \mathcal{E}} - e^{-\beta p \mathcal{E}}}{e^{\beta p \mathcal{E}} + e^{-\beta p \mathcal{E}}} = \tanh(\beta p \mathcal{E})
\end{equation}
and of induced dipole $\xi_i$ is
\begin{equation}
\left<\xi_i\right> = \frac{\int\limits_{-\infty}^{\infty} d\xi ~\xi e^{-\frac{\beta \xi^2}{2 \gamma} + \beta\mathcal{E}\xi} }{\sqrt{2\pi k_B T \gamma} ~ e^{\gamma \beta \mathcal{E}^2/2}} = \gamma \mathcal{E}
\end{equation}

Now, the electrostatic free energy can be written in the final form:
\begin{equation}
\label{el_free_en}
F_{el} = F_0 + N (\mathcal{E} - E) \eta(\mathcal{E}) + \frac{N \eta^2(\mathcal{E}) \chi}{2\epsilon_0},
\end{equation}
where in the thermodynamic limit $N\to \infty$ in the last term we have neglected the boundary corrections; $\chi$ is determined by the relation:
\begin{equation}
\chi = \frac{1}{4\pi}\sum\limits_{j(\neq i)=1}^N\frac{1}{r_{ij}^3}\left(1-3\cos^2 \theta_{ij}\right).
\end{equation}
The parameter $\chi$ describes the dipole-dipole interactions in the lattice. It can be treated as dipole analogue to the Madelung constant in an ionic solid. Also, the auxiliary function
\begin{equation}
\eta(\mathcal{E}) = \gamma \mathcal{E} + p S(\mathcal{E})
\end{equation}
has been introduced as the averaged total dipole moment of the node. Minimizing the total free energy (\ref{el_free_en}) with respect to $\mathcal{E}$, we obtain the self-consistent field equation:
\begin{equation}
E = \mathcal{E} + \frac{\chi}{\epsilon_0} \eta(\mathcal{E}). \label{eq.E}
\end{equation}
It is instructive to point out that in the case of small permanent dipole moment $p$ and weak electric fields $E$, Eq. (\ref{eq.E}) can be transformed into
\begin{equation}
\mathcal{E} = \frac{E}{1 + \frac{(\gamma + \beta p^2)\chi}{\epsilon_0}},
\end{equation}
which corresponds to the linear response theory limit.

Parameter $\kappa_0$ can be obtained from the condition of free energy minimum at $\alpha = \alpha_0$. As it follows from the derivation in Appendix II (see ESI), the reference state can be a state with initial stress. In the context of present consideration, the initial stress is not zero in the absence of external field ($E=0$), if the MOF at $\alpha = \alpha_0$ has nonzero spontaneous polarization, i.e. is in a ferroelectric state \cite{Landau.2013}. Following paper \cite{Ghoufi.2017}, we consider only that case. Thus, $\kappa_0$ is obtained from the extreme condition:
\begin{equation}
\kappa_0 = -\frac{\eta^2(\mathcal{E}_0)}{2\epsilon_0} \frac{d \chi(\alpha)}{d \alpha}\biggr\rvert_{\alpha = \alpha_0},
\end{equation}
where $\mathcal{E}_0$ is the solution to Eq. (\ref{eq.E}) with $E = 0$ and $\alpha = \alpha_0$.

Minimization of the total free energy with respect to the $\alpha$ will allow us to obtain the comprehensive description of the system of interest.

\section{Numerical results and discussions}
Here, we use the following set of dimensionless parameters: $\tilde p = p/(a^3 \epsilon_0 k_B T)^{1/2}$, $\tilde \chi = \chi a^3$, $\tilde \gamma = \gamma / \epsilon_0 a^3$, $\tilde E = E (\epsilon_0 a^3)^{1/2} (k_B T)^{-1/2}$, $\title A = A/a^2$. Let us discuss the behavior of the matrix unit cell area and the polarization density as a function of an external electric field.

\begin{figure}
\includegraphics[scale = 0.5]{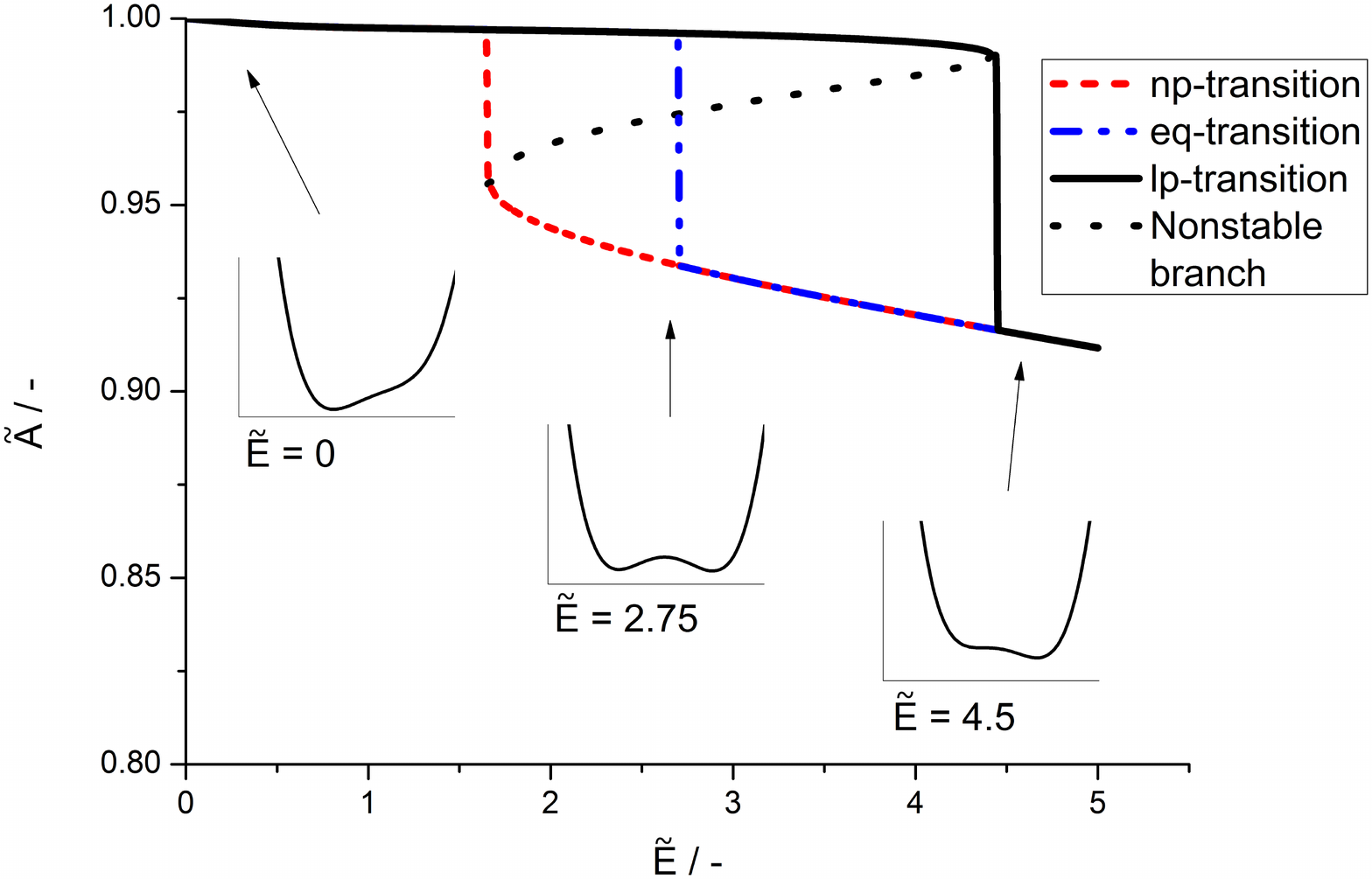}
  \caption{The cross-section area of the unit cell as a function of the an external electric field. The np-transition and lp-transition correspond to the situation when the matrix enters a stable state from metastable states corresponding to the narrow-pore and large-pore, respectively. The eq-transition corresponds to the equilibrium transition, i.e. to the intersection point of the free energies of lp- and np-phase. The non-stable branch corresponds to the maximum on the Helmholtz free energy. The insertions demonstrate the free energy as a function of angle ($\alpha$) at three different strength values of the external electric field ($\tilde{E}$ = 0, 2.75, 4.5).}
  \label{fig:1}
\end{figure}

At first, we discuss the influence of the external electric field on the cross-section area ($A$) of the MOF. The cross-section area is a function of the angle $\alpha$, so that we will discuss the changes in the context of the angle. Initially, the lattice exists in the state with the angle $\alpha_0$, we call this state - large pore ($lp$). After the application of the external electric field, the angle changes from its initial value to a bigger one (that in turn causes the decrease of $A$). Further field alteration generates abrupt change in $\alpha$ from $lp$ to narrow pore ($np$). Fig. \ref{fig:1} shows three possible types of transitions occurring in the 2D MOF. The np-transition and lp-transition correspond to the situation when the matrix enters a stable state from metastable states corresponding to the narrow-pore and large-pore, respectively. The eq-transition corresponds to the equilibrium transition, i.e. to the intersection point of the free energies of lp- and np-phase. In this case, the energy barrier between lp- and np-phases at the equilibrium transition point is very low. It is a possible sign that the obtained transition could be shifted or suppressed by other external stimuli, for example, gas adsorption or mechanical stress.

\begin{figure}
\includegraphics[scale = 0.5]{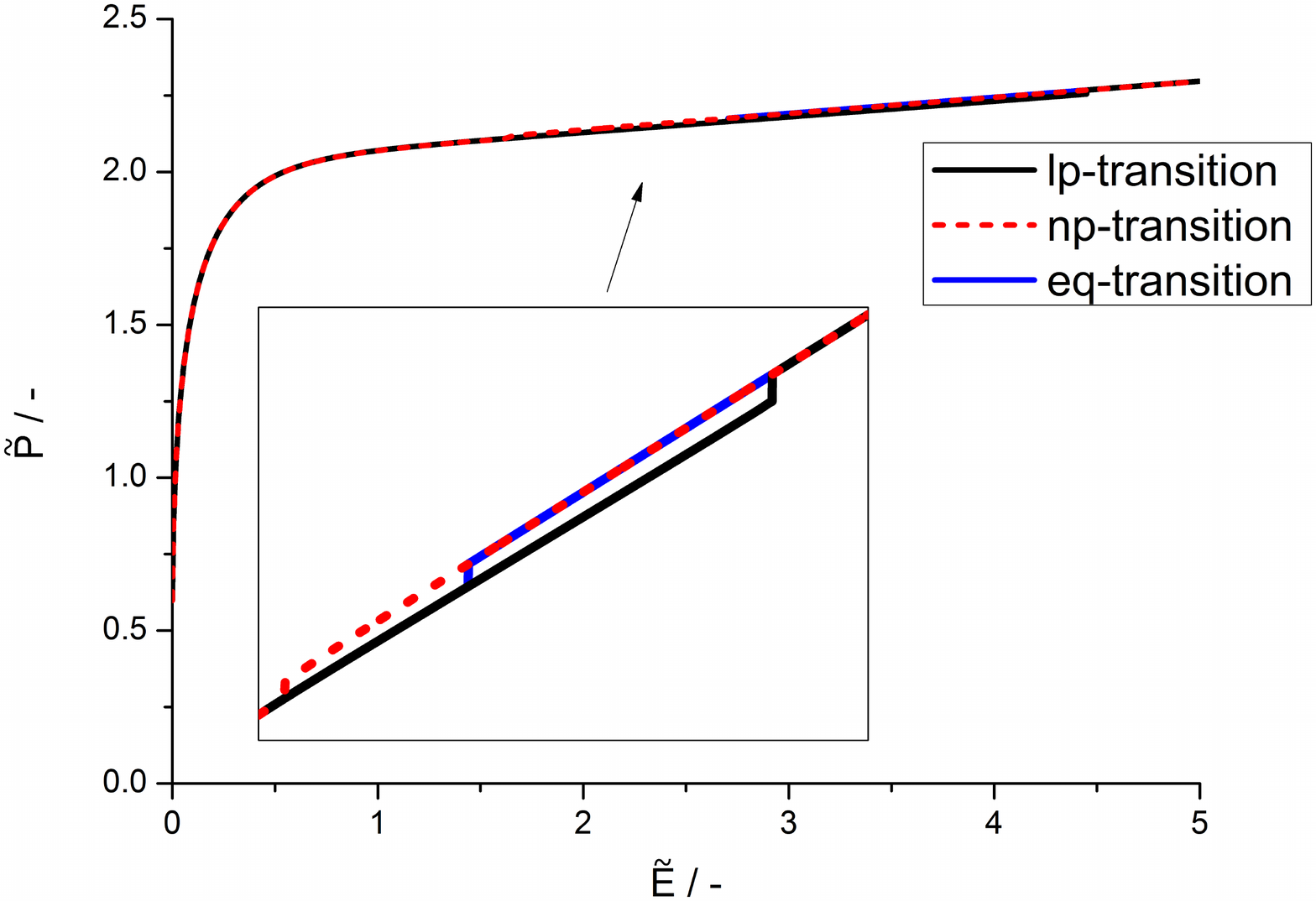}
  \caption{The polarization density, defined as total dipole moment of dielectric media divided per number of nodes, as a function of an external electric field. The insertion presents the magnified region corresponding to the structural transition in the matrix.}
  \label{fig:2}
\end{figure}

Fig. \ref{fig:2} shows the polarization density (the induced dipole moment of the dielectric divided by the number of nodes) as a function of an external electric field. There is an inflection point at $\tilde{E}\approx 1$, which corresponds to the saturation of orientation polarization. The following increase in polarization corresponds to the mechanism, based on molecular polarizability of the matrix nodes. The insertion shows the magnified region with hysteresis of the polarization curve. As in the case with the unit cell area, the three jumps correspond to two metastable transitions and one equilibrium transition.

The results from Fig. \ref{fig:1} are in qualitative agreement with MD simulation made by Ghoufi \cite{Ghoufi.2017}, where the authors obtained the hysteresis on the unit cell volume of the empty MIL-53(Cr) as a function of an external electric field. The polarization density obtained in our model demonstrates similar behavior at a low strength of the external electric field but does not contain a pronounced maximum in the region of structural transition. The latter may be the results of the model assumptions and simplifications e.g. the 2D dipoles model and not accounting for the real structure of the polarized unit.

The polarization process in this model is due to two effects: orientation polarizability of permanent dipoles and polarizability of induced dipoles. While the latter has no saturation, the permanent contribution can achieve it, when all permanent dipoles oriented along the external electric field. Thus, in the case of only permanent dipoles, the MOF phase transition can occur only at the values of the external electric field below the value corresponding to the saturation of the polarization density. The further increase of the field will not increase the contribution of the dipole-dipole interactions to the Helmholtz free energy and, thus, the system can be trapped in the large phase. On the other hand, if the host has only induced polarization mechanism, the phase transition should occur but at a higher value of electric field compared to the case when both contributions are included.

As mentioned in the Introduction, the structural transition occurs at the electric field of several GV/m. These values exceed the breakdown limit in the air, thus we can offer two possible situations when the field effect can be significant. The first one -- the MOF is already under other external stimuli and the electric field can shift the structural transition. The second one -- the material is significantly softer, then MIL-53(Cr) and there is no need in such high electric field strength to provoke the structural transition. Worth noting, that despite the "2D - assumption" the model should be applicable to the real materials with layered-like structures separated by the distances which is much bigger than the length of the linker in the plane. Also, the described theoretical formalism is, in general, applicable to 3D structures. However, that will require additional information about the material structure and accurate estimation of the stiffness constants. Such an analysis is beyond the scope of the present paper, while our main goal was to make a simple qualitative model of the electric field induced structural transition of a flexible metal-organic framework.

\section{Conclusions}
Herein, we report a simplified mathematical description of the structural transition of a metal-organic framework in an electric field based on a 2D lattice model. We have obtained the hysteresis on the unit cell cross-section area and on the polarization density as a function of the electric field. Despite the significant simplifications, our results are in qualitative agreement with the molecular dynamic simulation of similar phase transition in empty MIL-53(Cr)\cite{Ghoufi.2017}. Also, it is worth noting that our results confirm the qualitative picture proposed by Schmid\cite{Schmid.2017}. The availability of this model can provide the basis for future description and prediction of electric field induced phase transitions not only of MOFs but also of other materials.

\section{Supporting Information}
Supporting information consists of two Appendixes. In the first one, we show detailed calculations of $\chi$ parameter and in the second one - the derivation of elastic free energy.

\section{Appendix I: calculation of the dipole - dipole interaction parameter}
Here we present the calculation of the dipole - dipole interaction parameter $\chi$ which is defined as
\begin{equation}
   \chi = \frac{1}{4\pi}\sum\limits_{j(\neq i)=1}^N\frac{1}{r_{ij}^3}\left(1-3\cos^2 \theta_{ij}\right).
\end{equation}
Now, using the notations, introduced in the main text, we can rewrite it as the sum over $n$,$m$ indexes:
\begin{equation}
\nonumber
\chi = \frac{1}{8\pi a^3}\sum\limits_{n,m=1}^{\infty}
\frac{n^2\sin^2\frac{\alpha}{2}-2 m^2\cos^2\frac{\alpha}{2}}{\left(n^2\sin^2\frac{\alpha}{2}+m^2\cos^2\frac{\alpha}{2}\right)^{5/2}}
\end{equation}
\begin{equation}
\nonumber
+\frac{1}{8\pi a^3}\sum\limits_{n,m=0}^{\infty}
\frac{\left(n+\frac{1}{2}\right)^2\sin^2\frac{\alpha}{2}-2\left(m+\frac{1}{2}\right)^2\cos^2\frac{\alpha}{2}}{\left(\left(n+\frac{1}{2}\right)^2\sin^2\frac{\alpha}{2}+\left(m+\frac{1}{2}\right)^2\cos^2\frac{\alpha}{2}\right)^{5/2}}
\end{equation}
\begin{equation}
+\frac{1}{16\pi a^3\sin^3\frac{\alpha}{2}}\left(1-2\tan^3\frac{\alpha}{2}\right)\zeta(3),
\end{equation}
where $\zeta(s)=\sum\limits_{n=1}^{\infty}\frac{1}{n^s}$ is the zeta-function.

The first sum can be approximated as a double integral:
\begin{eqnarray}
\frac{1}{8\pi a^3}\sum\limits_{n,m=1}^{\infty}
\frac{n^2\sin^2\frac{\alpha}{2}-2 m^2\cos^2\frac{\alpha}{2}}{\left(n^2\sin^2\frac{\alpha}{2}+m^2\cos^2\frac{\alpha}{2}\right)^{5/2}} &\approx& \frac{1}{8\pi a^3} \int\limits_1^{\infty}dn\int\limits_1^{\infty}dm\frac{n^2\sin^2\frac{\alpha}{2}-2 m^2\cos^2\frac{\alpha}{2}}{\left(n^2\sin^2\frac{\alpha}{2}+m^2\cos^2\frac{\alpha}{2}\right)^{5/2}} = \nonumber\\
&=& \frac{1}{8\pi a^3 cos^3\frac{\alpha}{2}} \left[ \frac{1}{\sqrt{tan^2\frac{\alpha}{2}+1}} -  \frac{1}{tan\frac{\alpha}{2}} \right]
\end{eqnarray}

In the second sum we distinguished three contributions: from $n = 0$, from $m = 0$ and the remaining part of the sum. The sums corresponding to $n = 0$ and $m = 0$ are single and can be estimated with the Euler-Maclaurin formula (first or zero order), while the double sum is replaced with a double integral.

\begin{eqnarray}
\frac{1}{8\pi a^3}\sum\limits_{n,m=1}^{\infty}
\frac{\left(n+\frac{1}{2}\right)^2\sin^2\frac{\alpha}{2}-2\left(m+\frac{1}{2}\right)^2\cos^2\frac{\alpha}{2}}{\left(\left(n+\frac{1}{2}\right)^2\sin^2\frac{\alpha}{2}+\left(m+\frac{1}{2}\right)^2\cos^2\frac{\alpha}{2}\right)^{5/2}} \approx \nonumber\\
\frac{1}{12\pi a^3 cos^3\frac{\alpha}{2}} \left[ \frac{1}{\sqrt{tan^2\frac{\alpha}{2}+1}} -  \frac{1}{tan\frac{\alpha}{2}} \right].
\end{eqnarray}

The summation of the single sums ($n$,$m$ = 0) yields:
\begin{eqnarray}
\frac{1}{2 \pi a^3 cos^3\frac{\alpha}{2}} \left[ \frac{1}{\sqrt{tan^2\frac{\alpha}{2}+1}} -  \frac{1}{tan\frac{\alpha}{2}} \right].
\end{eqnarray}

Therefore, the final result for $\chi$ is:
\begin{eqnarray}
\chi = \frac{1}{8\pi a^3} \left[ \frac{17}{3\cos^2\frac{\alpha}{2}} \left(1 -  \csc\frac{\alpha}{2}\right) +\frac{\zeta(3)}{2\sin^3\frac{\alpha}{2}}\left(1-2\tan^3\frac{\alpha}{2}\right) \right]
\end{eqnarray}

\section{Appendix II: estimation of elastic energy}
In this appendix, we present the derivation of the elastic energy from the main text of the manuscript. From the general considerations of the theory of elasticity \cite{Landau.1986} the Helmholtz free energy of the solid body should be the function of strain tensor. Expanding the free energy density (with respect to the area in the reference state) in Taylor series and bearing in mind that the MOF deformations can be significant, we obtain:
\begin{eqnarray}
F_m &\approx& F_{m,0} + \sigma^{(0)}_{i_1i_2} u_{i_1i_2} + \frac{1}{2!}\lambda_{i_1..i_4} u_{i_1i_2} u_{i_3i_4} \\ \nonumber &+&\frac{1}{3!}\theta_{i_1..i_6} u_{i_1i_2} u_{i_3i_4} u_{i_5i_6} + \frac{1}{4!}\tau_{i_1..i_8} u_{i_1i_2} u_{i_3i_4} u_{i_5i_6} u_{i_7i_8},
\end{eqnarray}
where $F_{m,0}$ is the energy of the matrix in the reference state, $\boldsymbol{\sigma^{(0)}}$ is the initial stress in the material, $\boldsymbol{u}$ is the strain tensor, $\boldsymbol{\lambda}$ is the stiffness tensor \cite{Landau.1986} and $\boldsymbol{\theta}$, $\boldsymbol{\tau}$ are the expansion tensor coefficients determining the elastic behavior of the metal-organic framework beyond the linear elasticity. Further, following our approximation that the length of the linker between metal ions is constant under deformation, we propose the linear dependence of strain tensor on $\alpha$:
\begin{equation}
u_{ij} = a_{ij} (\alpha - \alpha_0),
\end{equation}
where $\alpha$ is the actual angle in the unit cell and $\alpha_0$ is the reference angle, $a_{ij}$ is the constant tensor independent of the angle. Thus, the elastic Helmholtz free energy can be written as:
\begin{equation}
F_m(\alpha)/N = F_{m,0} + \kappa_0 (\alpha - \alpha_0) + \kappa_1 (\alpha - \alpha_0)^2 + \kappa_2 (\alpha - \alpha_0)^3 + \kappa_3 (\alpha - \alpha_0)^4,
\end{equation}
where $\kappa_i$ are the "elastic" constants which are obtained as the convolutions of the above tensor coefficients with products of tensors $a_{ij}$ and defining the free energy profile with respect to the angle $\alpha$.

\section{Acknowledgments}
The reported study was funded by the RFBR according to research project No 18-31-20015.

\end{document}